# Possible Graviton Transitions and Gaser Action in High-$T_c$ Superconductors.

Giorgio Fontana

*University of Trento, Faculty of Science, 38050 Povo, TN, Italy*
*fontana@science.unitn.it*

It is well known that excited quantum systems can emit gravitons as well as photons. Differently from photon emission, spontaneous graviton emission is characterized by such a low probability that stimulated emission appears the only possible mechanism for producing gravitational radiation with quantum systems. The principles of the gravitational counterpart of the laser, which has been named gaser, have been theorized decades ago. Now a class of high temperature superconductors might be identified as a viable active material for practical gaser action.

1. INTRODUCTION

The study of the properties of gravitational radiation would be accomplished with much ease if a compact and efficient source of this radiation became available to experimenters.
Most of the studies on laboratory sources of gravitational radiation are based on rotating bodies [1] or mechanical vibrations in crystals or other suitable materials [2]. Other possible approaches are based on direct electromagnetic to gravitational field conversion [3].
With the exception of a near field experiment [1], none of them can offer a useful and easily detectable emission of gravitational radiation. This is a consequence of the weakness of gravity and the well known relations between the gravitational field and the other fields that are more easily produced in a known way [4], [5].
By including Einstein linearized field equations in quantum mechanics, it has been shown that transitions emitting gravitons do exist [6]. Compared to photon transitions, graviton transitions exhibit a very low probability, making the emission of gravitational radiation from quantum systems almost negligible, in agreement with the behavior of classical systems. On the other hand, stimulated emission is possible in quantum systems, and this effect has no analogy in classical systems. Stimulated emission of gravitons is the working principle of the gaser, therefore the gaser appears the most promising device for efficient production of gravitational radiation. In lasers, coherence is given by the photon field in a resonant cavity, in our gaser coherence will be provided by the collective wave-function of Cooper pairs.

## 2. PHOTON AND GRAVITON TRANSITIONS.

The theory describing the emission of gravitational radiation from quantum systems has been developed by Halpern and Laurent [6]. Transition for which the orbital quantum number *L* changes by ±2 and for which the total quantum number *J* changes by 0 or ±2 are gravitational quadrupolar transitions, for which the emission of photons is forbidden and the emission of gravitons is allowed. For comparison, the selection rules for photons are those for which the orbital quantum number *L* changes by ±1 and for which the total quantum number *J* changes by 0 or ±1, etc. For instance, for an atom, gravitational transitions are those between orbitals 3d and one characterized by a lower energy among 3s, 2s and 1s. It can be easily observed that in atomic systems gravitational transitions compete with multiple photon transitions, therefore methods for counteracting photon transitions have to be developed. Reference [6] reports about the possible use of superconducting mirrors around the active material for avoiding the emission of electromagnetic radiation, even if they are not considered perfect enough.

Detailed analysis comparing the emission of gravitons to the emission of photons has shown that the ratio *R* of the transition probabilities for matrix elements of equal structures is of the order of 1.6 $10^{-36}$ for the proton [6] and of the order of 4.8 $10^{-43}$ for the electron.

With *J=2*, the transition probability per unit of time for the emission of gravitons is proportional to:

$$M^2 (d \cdot f/c)^4 f \qquad\qquad 1)$$

This quantity increases with the square of the moment of inertia $Md^2$.

The frequency *f* is inversely proportional to the moment of inertia; thus in the quantum theory, the smaller the moment of inertia is, the higher the emission probability for gravitons becomes. Therefore, quantum systems composed by couples of closely interacting electrons seem favored for the emission of gravitons.

As already stated, it is certainly possible to increase further the emission of gravitational radiation by using induced emission at resonance. Because of the lack of mirrors or resonant cavities suitable for gravitational radiation, and taking into account the physical quantities involved, the characteristic length of the single pass molecular gaser has been found to be [6]:

$$l \approx 10^{38} \frac{\delta\omega}{\omega} \quad \text{cm,} \qquad\qquad 2)$$

where $\frac{\delta\omega}{\omega} = \Delta$ is the linewidth of the transition.

$\Delta=10^{-14}$ is a typical value for molecular transitions, therefore gaser action is apparently impossible to achieve with normal atomic systems and using acceptable lengths, nonetheless it might be interesting to describe the properties that the ideal gaser material should provide:

1. A quantum system with two energy levels characterized by a difference in orbital quantum number of 2.
2. A quantum system possibly composed of densely packed couples of closely interacting electrons.
3. A quantum system in which the two energy levels are respectively populated by objects with exactly the same wavefunction and the same energy in order to have negligible linewidths, thus permitting efficient stimulated emission.
4. A quantum system in which the two energy levels are populated by quantum objects, for which the two wavefunctions are orthogonal in order to prevent photon transitions and tunneling.

5.  A quantum system in which population inversion can be achieved to initiate and sustain gaser action.

A possible candidate material capable of fulfilling the enumerated requirements is a cuprate high temperature superconductor (HTSC) with orthorombic crystal structure, being the two quantum systems *s-wave* and *d-wave* Cooper-pairs.

3. GRAVITON TRANSITIONS IN HTSC AND ITS POSSIBLE OCCURRENCE.

The possibility of graviton transitions in HTSC might appear too speculative, instead the occurrence of the graviton transition seems to be indirectly shown in already published measurement data, we will simply propose an interpretation of the available experimental data that agree with our approach.
Superconductivity is characterized by a macroscopic quantum phase-coherent state. In Low-$T_c$ superconductors (Pb, Nb, $Nb_3Sn$, etc) the phonon-mediated electron-electron interaction produces spin-singlet pairing with *s-wave* symmetry (BCS) [7]. In High-$T_c$ cuprate superconductors the structure of Cooper pairs was extensively studied in recent years [8], [9], [10], [11], [12] with the conclusion that there is a predominant, but not exclusive, *d-wave* pairing.
Tunneling experiments using Josephson junctions between HTSC and Low-$T_c$ superconductors have consistently shown the presence of an admixture of *s-wave* and *d-wave* pairing in the order parameter of orthorombic cuprate HTSCs.
In the phenomenological Ginzburg-Landau theory of superconductivity, the order parameter is a wavefunction and after the formalization of the microscopic theory of superconductivity [7], Gor'kov [13] established the link between the old phenomenological and the new theory. Near $T_c$ the Ginzburg-Landau equations can be derived from the BCS theory and the order parameter is identified with the pair wave-function, more precisely it has been defined as the product of the wave-functions of all the pairs. Therefore, the admixture of *s-wave* and *d-wave* pairing in the order parameter seems to agree with the presence of a product of wave-functions of two populations of Cooper pairs, one associated to the *s-wave* component of the order parameter and the other associated to the *d-wave* component.
In [9] a best fit data analysis of a BSCCO sample named *mb232* is in agreement with the existence of these two populations with critical temperatures of 28K for $T_{c,d}$ and 10 K for $T_{c,s}$.
Because of Gor'kov result, the *s-wave* population is composed of Cooper pair with total spin S=0 in agreement with BCS theory. BCS theory suggests the possible existence of states with S≠0 with no energy gap respect to states with S=0, but measurements [9] show the existence of an additional state with S=2 and a measurable energy gap respect to the state with S=0 because of the different critical temperatures. The two states could be attributed to pairs confined in the CuO layers and to pairs not confined in the layers. This is not in contradiction with BCS. In fact, in anisotropic HTSC two pairing states have been shown to exist, but only the state with lower energy is stable, the other state must be empty and it manifests its existence only in experiments that create a direct coupling to it with wave-function selective Josephson-junctions.
Because pairs in the two states differ for the total energy and the quantized angular momentum, we now relate the measured properties of mixed order parameter in HTSCs to the possible graviton transitions. The difference of critical temperatures is a measure of the energy gap. For the energy gap Δ, BCS predicts:

$$\Delta(T) \approx 1.76 k_B T_c \sqrt{\cos\left[\frac{\pi}{2}\left(\frac{T}{T_c}\right)^2\right]} \qquad 3)$$

For T< $T_{c,d}$ /2 and T< $T_{c,s}$ /2, the term under the square root can be neglected, giving for the cited sample a differential *s-wave/d-wave* energy gap of 2.74 $10^{-3}$ eV and a line frequency for the gravitational transition of 660 GHz. On the other hand the coefficient 1.76 in equation 3 has been measured to be as large as 3.6 for BSCCO, this may increase the frequency up to about 1.3 THz. The power emitted in the form of gravitational radiation is formally related to the recombinantion rate, corrected for the momentum conserving reaction on the crystal lattice, which will be converted to heat.

The condensation of both the *s-wave* component and the *d-wave* component ensures that the relative energies of the pairs are almost not affected by thermal fluctuations and impurities except for distances less than the coherence length. On the other hand, the linewidth is critical for achieving amplification, but no experimental data is available on the residual relative fluctuations of the pair energies.

The strong correlation among pairs is the main difference between superconductivity and other forms of Bose-condensation such as superfluidity and Bose-condensates of trapped atoms [14]. Because of the strong correlation, fluctuations are averaged over the pairs. In fact, the Josephson effect [15] and the large literature on SQUIDs indicate that the coherence of the wave-function is maintained over very large distances, with no apparent local splitting of the ground state into fine structures or the appearance of drifts, which is also prohibited by BCS theory itself.

Being the *d-wave* component the one with the highest binding energy, it represents the state with the highest population. To achieve gaser action it is therefore necessary to increase the density of the metastable *s-wave* component.

A possible mechanism for achieving population inversion could be the electrical injection of *s-wave* Cooper pairs with a Josephson junction made with Low-$T_c$ materials, which couple only to the *s-wave* component of the HTSC. The *s+d* component can be removed by a normal metal contact. The transport is performed in the bulk superconductor and it requires no work; power is required for the removal of *s+d* Cooper pairs and for the creation of *s* Cooper pairs in the Low-$T_c$ material.

Reference [9] presents detailed experimental data. It has been found that in a Pb/BSCCO Josephson junction, $T_{c,Pb}$ is different from, $T_{c,s}$ in BSCCO, with a strong indication that the *s-wave* component in BSCCO is not induced by proximity to Pb. In addition, detailed discussion of experimental results [9] indicates that second order tunneling [16] may not play a substantial role in the tunneling process: the *s-wave* component is present in BSCCO, but only in the vicinity of the junction. This behavior might be interpreted as a very short diffusion length and consequently a highly efficient spin-2 recombination.

So far, the observed Josephson tunneling between BSCCO and Pb has suggested two interpretations, the first one is the homogeneous presence of a mixed order parameter in orthorombic HTSCs, the second one is the homogeneous presence of a mixed order parameter in Pb [17]. A third possibility is the one presented here, in our model a coherent gravitational radiation field allows *s-wave* tunneling to an available energy level of the Cooper pair condensate in BSCCO, followed by a rapid transition of the pairs to the dominant *d-wave* symmetry; being the low probability of the process compensated by the smallness of the linewidth.

The emission of gravitational radiation could explain the observed behaviour of Pb/BSCCO Josephson junctions [9], especially the fact that the degree of admixture of the order parameter can dynamically depend on the mentioned processes.

As reported in the introduction, multiple photon transitions and multipole electromagnetic transitions compete with graviton transitions. Therefore it must be established if electromagnetic radiation can escape or, on the contrary, it is trapped in the bulk superconductor as required by the basic gaser theory [6].

Reported values of the surface resistance of HTSCs at THz frequencies [18, 19] seems to indicate that if multipole electromagnetic transitions could take place in the bulk superconductor, the associated electromagnetic radiation could not escape the material except for transitions localized not more than few hundred nanometers below the surface. References report a surface resistance of YBCO thin films at 77K – 1.3Thz of 1Ω, that is about an order of magnitude higher than that of gold at the same temperature, penetration depths of the order of 200nm have been also measured for YBCO.

4. CONCLUSION.

We have discussed the possibility of including graviton transitions in the model of high temperature superconductors. Graviton transitions between *s-wave* and *d-wave* Cooper-pairs might explain the puzzling behaviour of a class of BSCCO-LTSC Josephson-junctions.

Moreover, it is well known that quantum coherence in Bose-condensates would be the source of coherent radiation. For instance, the stimulated scattering of indirect excitons in coupled quantum wells has been attributed to a degenerate Bose-gas of electron-hole excitons at ultra-cryogenic temperatures in GaAs [20]. Stimulated emission of atoms in a Bose-condensate of electromagnetically trapped atoms has been also demonstrated [21].

The stimulated emission of gravitons from mixed *s-wave* / *d-wave* highly saturated Bose-condensates seems therefore an interesting possibility. Unfortunately, the model of HTSCs is still under discussion and no rigorous formal derivation of a HTSC gaser theory have been made. Nevertheless a possibility seems to exist that gaser action could take place in HTSC, provided that the interpretation of available references turns out to be consistent with the presented picture.

Considering the available data, experimental investigation seems more appropriate than the development of a full theory in order to prove or disprove the possibility that HTSCs could produce coherent high frequency gravitational radiation.

---

♦ also available at: http://www.ing.unitn.it/~fontana/halplaur.html

18) I. Wilke, M. Khazan and C. T. Rieck, *Terahertz surface resistance of high temperature superconducting thin films*, Journal of Applied Physics, Vol. 87, No. 6, 15 March 2000, 2984-2988.

19) Max Khazan, Ingrid Wilke and Christopher Stevens, *Surface impedance of Tl-2212 thin films at THz-frequencies*, IEEE Transactions on Applied Superconductivity, Vol 11, pt III, 2001, pp 3537-3540.

20) L.V. Buotv. A. L. Ivanov, A. Imamoglu, P. B. Littlewood, A. A. Shashkin, V. T. Dolgopolov, K. L. Kampman, A. C. Gossard, *Stimulated Scattering of Indirect Excitons in Coupled Quantum Wells: Signature of a Degenerate Bose-Gas of Excitons,* Physical Review Letters -- June 11, 2001 -- Volume 86, Issue 24, pp. 5608-5611

21) M. O. Mewes, M. R. Andrews, D. M. Kurn, D. S. Durfee, C. G. Townsend, and W. Ketterle, *Output Coupler for Bose-Einstein Condensed Atoms*, Physical Review Letters, Vol. 78, No. 4, 27 January 1997, pp. 582-585.